\documentclass{llncs}

\usepackage{amsmath}
\usepackage{url}
\usepackage{amssymb}
\usepackage{amsfonts}
\usepackage[british]{babel}
\usepackage{graphicx}
\usepackage{subfigure}
\usepackage{algorithm}
\usepackage[noend]{algorithmic}
\usepackage{a4wide}
\usepackage{color}

\newcommand{\cX}{{\mathcal X}}
\newcommand{\cC}{{\mathcal C}}
\newcommand{\cT}{{\mathcal T}}
\newcommand{\cD}{{\mathcal D}}
\newcommand{\cR}{{\mathcal R}}

\newtheorem{observation}{Observation}

\begin{document}

\title{When Two Trees Go to War}
\author{Leo van Iersel\inst{1}\thanks{Leo van Iersel was funded by the Allan Wilson Centre for Molecular Ecology and Evolution.}, Steven Kelk\inst{2}\thanks{Steven Kelk was 
funded by a Computational Life Sciences grant of The Netherlands Organisation for Scientic Research (NWO).}}
\institute{University of Canterbury, Department of Mathematics and Statistics,\\Private Bag 4800, Christchurch, New Zealand. l.j.j.v.iersel@gmail.com \\ \and Centrum voor Wiskunde en Informatica (CWI)\\ P.O. Box 94079,
1090 GB Amsterdam, The Netherlands. s.m.kelk@cwi.nl}
\maketitle

\begin{abstract}
Rooted phylogenetic networks are often constructed by combining trees, clusters, triplets or characters into a single network that in some well-defined sense simultaneously represents them all. We review these four models and investigate how they are related. In general, the model chosen influences the minimum number of reticulation events required. However, when one obtains the input data from two binary trees, we show that the minimum number of reticulations is independent of the model. The number of reticulations necessary to represent the trees, triplets, clusters (in the softwired sense) and characters (with unrestricted multiple crossover recombination) are all equal. Furthermore, we show that these results also hold when not the number of reticulations but the level of the constructed network is minimised. We use these unification results to settle several complexity questions that have been open in the field for some time. We also give explicit examples to show that already for data obtained from three binary trees the models begin to diverge.
\end{abstract}

\section{Introduction}
Consider a set of taxa~$\cX$. A \emph{rooted phylogenetic network} on~$\cX$ is a rooted directed acyclic graph in which the outdegree-zero nodes (the \emph{leaves}) are bijectively labelled
by~$\cX$. It is common to identify a leaf with the taxon it is labelled by and it is usually assumed that there are no nodes with indegree and outdegree one; we adopt both conventions.
Nodes with indegree at least two are called \emph{reticulations}. The edges entering a reticulation are called \emph{reticulation edges}. Nodes that are not reticulations are called \emph{tree nodes}. A phylogenetic network is called \emph{binary} if all reticulations have indegree two and outdegree one and all other nodes have outdegree zero or two.

One of the main challenges in phylogenetics is to reconstruct phylogenetic networks from biological data of currently living organisms. The reticulations in a phylogenetic network are of
special biological interest. These nodes represent ``reticulate'' evolutionary phenomena like hybridisation, recombination or lateral (horizontal) gene transfer. Motivated by the parsimony
principle, a phylogenetic network with fewer reticulations is often preferred over a network with more reticulations, when both networks represent the available data equally well.

Thus, we define the following fundamental problem \textsc{MinRet}. Given some set~$\cD$ of data describing some set~$\cX$ of taxa, find a phylogenetic network on~$\cX$ that ``represents'' $\cD$ and contains a minimum number of reticulations over all phylogenetic networks on~$\cX$ representing~$\cD$. We consider three specific variants of this problem: \textsc{MinRetTrees}, \textsc{MinRetTriplets} and \textsc{MinRetClusters}, for data~$\cD$ consisting of trees, triplets and clusters respectively.

The following subtlety has to be taken into account when reticulations with indegree higher than two are considered. When counting such reticulations, indegree-$d$ reticulations are counted $d-1$ times, because such reticulations represent $d-1$ reticulate evolutionary events (of which the order is not specified). Hence, using~$\delta^-(v)$ to denote the indegree of a node~$v$, we formally define the \emph{number of reticulations} in a phylogenetic network~$N=(V,E)$ as
\[\sum_{\substack{v\in V: \delta^-(v)>0}}(\delta^-(v)-1) = |E| - |V| + 1 \enspace.\]

\begin{figure}[h]
  \centering
  \includegraphics[scale=.5]{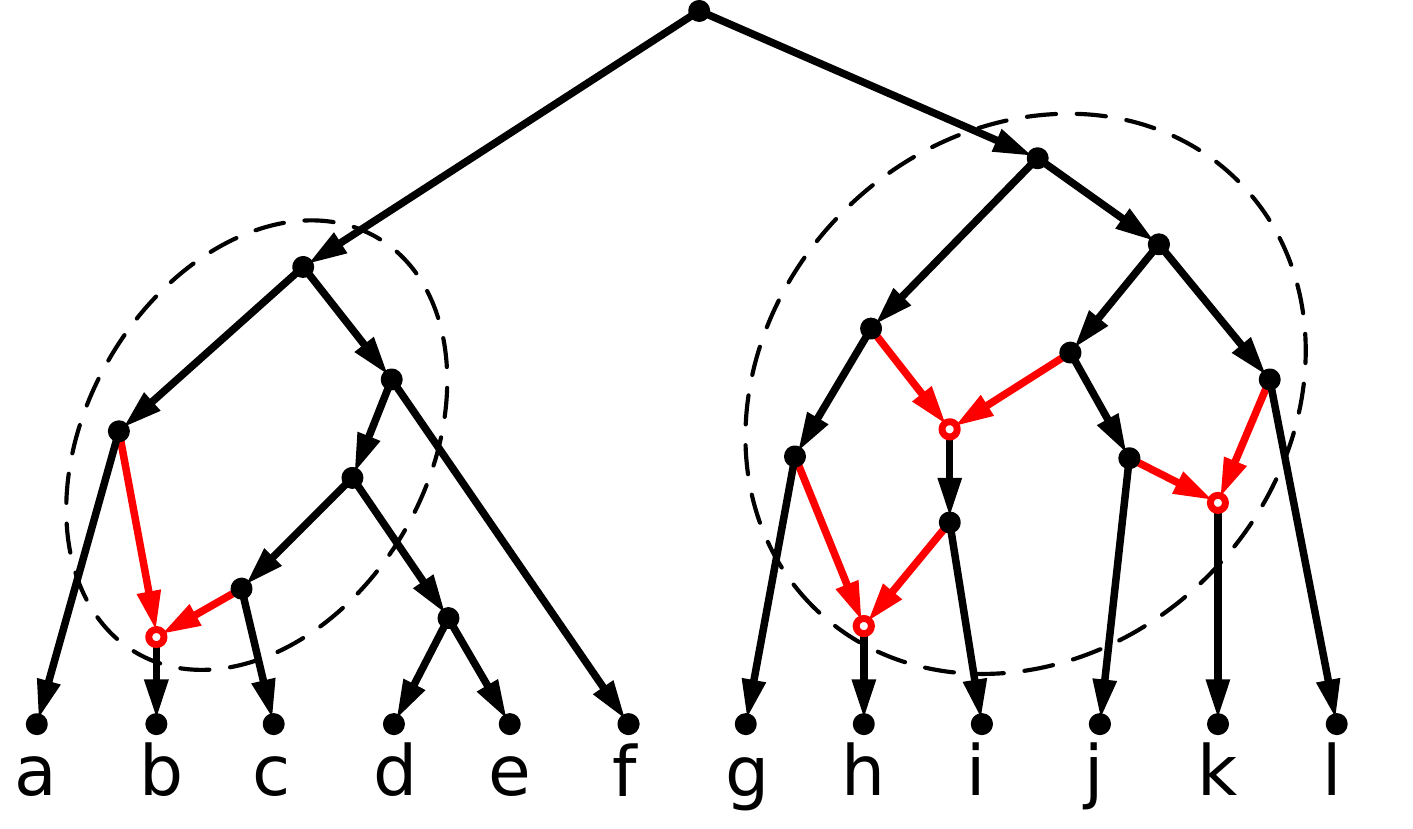}
  \caption{A level-3 phylogenetic network with four reticulations. Nontrivial biconnected components are encircled by dashed lines. Reticulations are unfilled and colored red. Reticulation edges are also indicated in red.}
  \label{fig:bcc}
\end{figure}

Instead of minimizing the total number of reticulations in a network, another possibility is to minimize the number of reticulations in each nontrivial biconnected component 
(informally: tangled part) of a network. Formally, a \emph{biconnected component} is a maximal subgraph that cannot be disconnected by removing a single node. A biconnected 
component
is \emph{trivial} if it is equal to a single edge and \emph{nontrivial} otherwise.
 For~$k\in\mathbb{N}$, a 
phylogenetic network is called a \emph{level}-$k$
network if each nontrivial biconnected component contains at most~$k$ reticulations. See Figure~\ref{fig:bcc} for an example of a phylogenetic network with four reticulations. 
This is a level-3 network, because each nontrivial biconnected component contains at most three reticulations.

This leads to the definition of the following \textsc{MinLev} variant of the fundamental problem. Given some set~$\cD$ of data describing some set~$\cX$ of taxa, find a level-$k$ phylogenetic network that ``represents'' $\cD$ such that~$k$ is as small as possible. There are again three versions: \textsc{MinLevTrees}, \textsc{MinLevTriplets} and \textsc{MinLevClusters}, for data~$\cD$ consisting of trees, triplets and clusters respectively.

The definition of ``represents'' heavily depends on the nature of the data in~$\cD$. We will discuss four types of data: trees, triplets, clusters and binary characters. 
Throughout the paper we assume a fixed set~$\cX$ of taxa.

\subsection{Trees}
\label{subsec:treedefs}
A \emph{phylogenetic tree} on~$\cX$ is a phylogenetic network on~$\cX$ without reticulations. There exist numerous methods that construct phylogenetic trees, for example from DNA data. These methods include Maximum Likelihood, Maximum Parsimony, Bayesian- and distance-based methods like Neighbor Joining. When phylogenetic trees are constructed for several parts of the genome separately (e.g. several genes), one often obtains a number of different phylogenetic trees. The same can occur when several phylogenetic trees are constructed using different methods.

\begin{figure}[t]
  \centering
  \includegraphics[scale=.5]{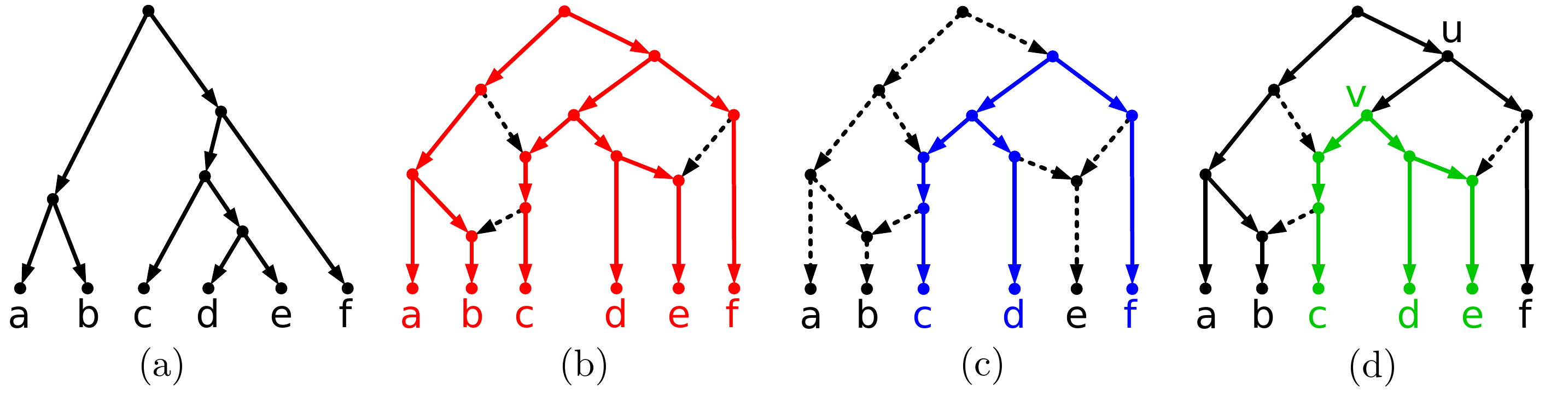}
  \caption{A phylogenetic tree~$T$ (a) and a phylogenetic network~$N$ (b,c,d); (b) illustrates in red that~$N$ displays~$T$ (edges not in the subdivision are dashed); (c) illustrates in blue that~$N$ is consistent with the triplet $cd|f$ from~$T$ (edges not in the embedding are again dashed); (d) illustrates in green that~$N$ represents cluster $\{c,d,e\}$ from~$T$ in the softwired sense (dashed reticulation edges are ``switched off'').}
  \label{fig:treestripletsclusters}
\end{figure}

Thus, given a number of phylogenetic trees, it is interesting to find a phylogenetic network that ``represents'' each of them. This is formalized by the notion of ``display'' as follows. A phylogenetic tree~$T$ is \emph{displayed} by a phylogenetic network~$N$ if~$T$ can be obtained from some subtree of~$N$ by suppressing nodes with indegree one and outdegree one (i.e. if some subtree of~$N$ is a subdivision of~$T$). See Figure~\ref{fig:treestripletsclusters} for an example.

\noindent For a set~$\cT$ of phylogenetic trees on~$\cX$, we define:
\begin{itemize}
\item[$\bullet$] $r_t(\cT)$ as the minimum number of reticulations in any phylogenetic network on~$\cX$ that displays each tree in~$\cT$ and
\item[$\bullet$] $\ell_t(\cT)$ as the minimum~$k$ such that there exists a level-$k$ phylogenetic network on~$\cX$ that displays each tree in~$\cT$.
\end{itemize}

The computation of $r_t$ has received much attention in the literature. For two binary
trees on the same taxon set the problem is NP-hard and APX-hard \cite{bordewich} although on the positive side it is
fixed-parameter tractable in $r_t$ \cite{sempbordfpt2007}\cite{bordewich2}; \cite{Semple2007} offers a good overview
of these and related results. These algorithmic insights have been translated into the software \textsc{HybridNumber} \cite{bordewich2} and its more advanced successor \textsc{HybridInterleave} \cite{quantifyingreticulation}. These programs compute $r_t$ exactly for two binary trees on the same taxon set. The program \textsc{SPRDist} \cite{wuISBRA2010} solves the same problem (using integer linear programming) and the program \textsc{PIRN} \cite{pirnISMB2010} can compute lower and upper bounds on $r_t$ for any number of binary trees on the same taxon set. In \cite{huynh} a polynomial-time algorithm is described that constructs a level-1 phylogenetic network that displays all trees and has a minimum number of reticulations, if such a
network exists.

\subsection{Triplets}
\label{subsec:tripdefs}

A \emph{(rooted) triplet} on~$\cX$ is a binary phylogenetic tree on a size-3 subset of~$\cX$. We use $xy|z$ to denote the triplet with taxa~$x,y$ on one side of the
root and $z$ on the other side of the root. Triplets can be constructed using any of the methods for constructing phylogenetic trees (using a fourth taxon as an
outgroup in order to root the triplet). Alternatively, one can first construct one or more phylogenetic trees and subsequently find the set of triplets that are
contained in these trees. The main motivation for the latter approach is that representing all triplets might require fewer reticulations than representing the entire
trees.

This can be formalised by using the notion of \emph{display} introduced above. For triplets, often ``consistent with'' is used instead of ``displayed by''. A triplet~$xy|z$ is \emph{consistent} with a phylogenetic network~$N$ (and~$N$ is \emph{consistent} with~$xy|z$) if~$xy|z$ is displayed by~$N$. See Figure~\ref{fig:treestripletsclusters} for an example. Given a phylogenetic tree~$T$ on~$\cX$, we let~$Tr(T)$ denote the set of all rooted triplets on~$\cX$ that are consistent with~$T$. For a set of phylogenetic trees~$\mathcal{T}$, we let~$Tr(\mathcal{T})$ denote the set of all rooted triplets that are consistent with some tree in~$\mathcal{T}$, i.e. $Tr(\mathcal{T})=\bigcup_{T\in\mathcal{T}}Tr(T)$.

\noindent For a set~$\cR$ of triplets on~$\cX$, we define:
\begin{itemize}
\item[$\bullet$] $r_{tr}(\cR)$ as the minimum number of reticulations in any phylogenetic network on~$\cX$ that is consistent with each triplet in~$\cR$ and
\item[$\bullet$] $\ell_{tr}(\cR)$ as the minimum~$k$ such that there exists a level-$k$ phylogenetic network on~$\cX$ that is consistent with each triplet in~$\cR$.
\end{itemize}

Throughout the article we will write $r_{tr}(\mathcal{T})$ and $\ell_{tr}(\mathcal{T})$ as abbreviations
for $r_{tr}(Tr(\mathcal{T}))$ and $\ell_{tr}(Tr(\mathcal{T}))$ respectively.

A triplet set $\mathcal{R}$ on $\mathcal{X}$ is said to be \emph{dense} when, for every three distinct
taxa $x,y,z \in \mathcal{X}$, at least one of $xy|z, xz|y, yz|x$ is in $\mathcal{R}$ \cite{JanssonEtAl2006}.
Given a dense triplet set, \cite{JanssonEtAl2006}\cite{JanssonSung2006} describe a polynomial-time algorithm that constructs a
level-1 network displaying all triplets, if such a network exists. The algorithm \cite{simplicityAlgorithmica} can be used to find such a network that also minimizes the
number of reticulations, and this is available as the program \textsc{Marlon} \cite{MARLON}. These results have later been extended to
level-2~\cite{lev2TCBB}\cite{simplicityAlgorithmica}
(see also the program \textsc{Level2} \cite{LEV2program}) and more recently to level-$k$, for all~$k\in\mathbb{N}$~\cite{ToHabib2009}. The program \textsc{Simplistic}
\cite{SIM08}\cite{simplicityAlgorithmica} can be used to construct (simple) networks of arbitrary level (again, assuming density).

\subsection{Clusters}
\label{subsec:clusters}

A \emph{cluster} on~$\cX$ is a proper subset of~$\cX$. Clusters can be obtained from morphological data (e.g. species
with wings, species with eight legs, etc.) or from phylogenetic trees. The latter approach has a similar motivation as
in triplet methods. The clusters from the trees might be representable using fewer reticulations than that would be
necessary to represent the trees themselves. In addition, the clusters described by a phylogenetic tree are
biologically the most interesting features of the tree, because they describe putative monophyletic groups of species
(clades).

We use~$Cl(T)$ to denote the set of clusters of a phylogenetic tree~$T$, i.e. for each edge~$(u,v)$ of~$T$, the
set~$Cl(T)$ contains a cluster consisting of those taxa that are reachable by a directed path from~$v$. For a
set~$\cT$ of phylogenetic trees, we define $Cl(\cT)=\bigcup_{T\in\cT} Cl(T)$.

Similar to tree- and triplet methods, the general aim of cluster methods is to construct a phylogenetic network that
``represents'' some set of input clusters. There are two different notions of ``representing'' for clusters: the
``hardwired'' and the ``softwired'' sense. Given a cluster~$C\subset\cX$ and a phylogenetic network~$N$ on~$\cX$, we
say that~$N$ \emph{represents} $C$ \emph{in the hardwired sense} if there exists an edge~$(u,v)$ in~$N$ such that~$C$ is
the set of taxa reachable from~$v$ by a directed path \cite{HusonRupp2008}.

The definition of ``representing'' in the ``softwired sense'' is longer but biologically more relevant. We say that~$N$ \emph{represents} $C$ \emph{in the softwired sense} if there exists an edge~$(u,v)$ in~$N$ such that~$C$ is the set of taxa reachable from~$v$ by a directed path, when for each reticulation~$r$ exactly one its incoming edges is ``switched on'' and all other edges entering~$r$ are ``switched off'' (see Figure~\ref{fig:treestripletsclusters}). As a direct consequence,~$C$ is represented by~$N$ in the softwired sense if and only if there exists a phylogenetic tree~$T$ on~$\cX$ that is displayed by~$N$ and has~$C\in Cl(T)$. In this article, we do not consider cluster representation in the hardwired sense and therefore often write ``represents'' as short for ``represents in the softwired sense''.

\noindent For a set of clusters~$\cC$ on~$\cX$, we define:
\begin{itemize}
\item[$\bullet$] $r_c(\cC)$ as the minimum number of reticulations in any phylogenetic network on~$\cX$ that represents all clusters in~$\cC$ in the softwired sense and
\item[$\bullet$] $\ell_c(\cC)$ as the minimum~$k$ such that there exists a level-$k$ phylogenetic network on~$\cX$ that represents all clusters in~$\cC$ in the softwired sense.
\end{itemize}

We write $r_c(\cT)$ as shorthand for $r_c(Cl(\cT))$ and $\ell_c(\cT)$ as shorthand for $\ell_c(Cl(\cT))$.

A network is a \emph{galled network} if it contains no path between two reticulations that is contained in a single biconnected component.
In \cite{HusonKloepper2007} and \cite{husonetalgalled2009} an algorithm is described for constructing a galled network
representing~$\mathcal{C}$ in the softwired sense. In \cite{cassISMB} the algorithm \textsc{Cass}~\cite{cassdownload} is presented which aims at constructing a low-level network that represents~$\mathcal{C}$. \textsc{Cass} always returns a network representing all input clusters and, when $\ell_{c}(\cC)\leq 2$, it is guaranteed to compute $\ell_{c}$ exactly. Alongside the algorithms from \cite{husonetalgalled2009}\cite{HusonKloepper2007}\cite{HusonRupp2008}
\textsc{Cass} is available as part of the program \textsc{Dendroscope} \cite{Huson2007Dendroscope}.

\subsection{Binary character data}

Within the field of population genomics the literature on phylogenetic networks has evolved along a slightly different
route to the literature on trees, triplets and clusters. At the level of populations the principle reticulation event
is the \emph{recombination}, and in this context phylogenetic networks are sometimes called \emph{recombination networks}.
To avoid repetition we refer to \cite{gusfield2}\cite{gusfielddecomp2007}\cite{WuG08}
for background and definitions. In this article we will always assume that
recombination networks are constructed from \emph{binary} character data and that the root sequence is the all-0 sequence i.e. we are
dealing with the ``root known'' variant of the problem. We assume thus that the input is a binary $n \times m$ matrix $M$.

The basic definition given in \cite{gusfield2} is for the \emph{unrestricted multiple crossover} variant of the recombination
network model. Stated informally this means that, at each reticulation, each character can freely ``choose''
from which of its parents it inherits its value. This is quite different to the \emph{single crossover} variant
which has received far more attention in the literature. In the single crossover variant the sequence at a reticulation
is forced to obtain a prefix from one of its parents, and a suffix from the other, thus modelling chromosomal crossover.

\noindent For a binary matrix $M$, we define:
\begin{itemize}
\item[$\bullet$] $r_{sc}(M)$ as the minimum number of reticulations required by a recombination network that represents $M$, assuming the single crossover variant and an all-0 root, and
\item[$\bullet$] $r_{uc}(M)$ as the minimum number of reticulations required by a recombination network that represents $M$, assuming the unrestrained multiple crossover variant and an
all-0 root.
\end{itemize}
Given that the latter is a relaxation of the former, it
is immediately clear that for any input $M$,
\begin{equation}
r_{uc}(M) \leq r_{sc}(M).
\label{eq:scuc}
\end{equation}
In \cite{wang} it was claimed that it is NP-hard to compute $r_{uc}$. However,
\cite{bordewich} subsequently discovered that the proof in \cite{wang} was partially incorrect and modified it to prove that
computation of $r_{sc}$ is NP-hard.

There are some definitional subtleties when trying to map between the model of \cite{gusfield2} and the other models summarised in this
article. Some differences between the models are rather arbitrary and minor and thus easy to overcome,
and we do not discuss them here. In this article we restrict ourself to a more fundamental comparison concerning (under an appropriate transformation) the values
$r_{sc}$, $r_{uc}$ and $r_{c}$.\\
\\
The problem of computing $r_{sc}$ (in defiance of its NP-hardness) has attracted much attention. Articles such as
\cite{gusfield2}\cite{gusfielddecomp2007}\cite{WuG08}\cite{song2}\cite{lyngs2005} give a good overview of the methods in use. Much energy has
been invested in computing lower bounds for $r_{sc}$ (e.g. the program \textsc{HapBound} \cite{song2}), and some lower bounding techniques
also produce valid lower bounds for $r_{uc}$ (e.g. \cite{gusfield2}). Programs such as \textsc{Shrub} \cite{song2} produce upper bounds on
$r_{sc}$, and \textsc{Beagle} \cite{lyngs2005} uses integer linear programming to compute $r_{sc}$ exactly (for small instances). The programs
\textsc{HapBound-GC} and \textsc{Shrub-GC} compute lower and upper bounds on a value that lies somewhere between $r_{sc}$ and $r_{uc}$ \cite{crossover}. As
in other areas of the phylogenetic network literature the problem of computing $r_{sc}$ in a topologically constrained space of networks \cite{GusfieldEtAl2004}
has also been considered.

\subsection{Summary of Results}

In this article, we study how several methods for constructing phylogenetic networks are related. We begin by clarifying the relationship between phylogenetic networks that represent
clusters in the softwired sense and recombination networks that represent binary character data. We explain that the two models are equivalent when unrestricted multiple crossover
recombination is considered but fundamentally different when single crossover recombination is used. This clarification is necessary to place the main results from this article in the correct context.

We then turn to the problem of constructing phylogenetic networks from trees, triplets or clusters. In particular, we focus on triplets and clusters obtained from a set of 
trees on the same set of taxa. We show that the number of reticulations required to display the triplets is always less than or equal to the number of reticulations necessary 
to 
represent all clusters, and the latter number is in turn less than or equal to the number of reticulations necessary to display the trees themselves:
$$r_{tr}(\mathcal{T}) \leq r_{c}(\mathcal{T}) \leq r_{t}(\mathcal{T})\enspace .$$

We give examples for which these inequalities are strict i.e. an example in which the triplets need strictly fewer reticulations than the clusters and an example in which the clusters
need strictly fewer reticulations than the trees.

However, the main result of this article shows that, when one considers a set~$\cT$ containing two binary trees on the same set of taxa, the numbers of reticulations required to represent the triplets, clusters or the trees themselves are all equal:
$$r_{tr}(\mathcal{T}) = r_{c}(\mathcal{T}) = r_{t}(\mathcal{T})\enspace .$$

In addition, all the results above also hold for minimizing level. In particular:
$$\ell_{tr}(\mathcal{T}) = \ell_{c}(\mathcal{T}) = \ell_{l}(\mathcal{T})\enspace .$$

These unification results turn out to have important consequences. We use the equalities above to settle several complexity questions that have been open for some time and
to strengthen several existing complexity results. In particular, we show that computation of $\ell_{t}(\mathcal{T})$, $r_{c}(\mathcal{T})$, $\ell_{c}(\mathcal{T})$, $r_{tr}(\mathcal{T})$
and $\ell_{tr}(\mathcal{T})$ are all NP-hard and APX-hard even when~$\cT$ consists of two binary trees on the same set of taxa. Thus, problems \textsc{MinRetTriplets}, \textsc{MinRetClusters}, \textsc{MinLevTrees}, \textsc{MinLevTriplets} and \textsc{MinLevClusters} are all NP-hard and APX-hard.

\section{Spot the difference}
\label{sec:spotdiff}

\subsection{Clusters and binary character data}

We say that two clusters $C_1, C_2 \subset \cX$ are \emph{compatible} if either $C_1 \cap C_2 = \emptyset$ or $C_1 \subset C_2$ or $C_2 \subset C_1$ and \emph{incompatible} otherwise.

Let $\mathcal{C}$ be a set of clusters on $\mathcal{X}$. Let $\mathcal{X} = \{x_1, ..., x_n\}$ and $\mathcal{C} = \{c_1, ..., c_m\}$ i.e. impose an arbitrary ordering on $\mathcal{X}$ and $\mathcal{C}$. The \emph{matrix encoding} of $\mathcal{C}$ is a binary matrix $Mat(\mathcal{C})$ with $n$ rows and $m$ columns. $Mat(\mathcal{C})_{i,j}$ has the value 1 if and only if $c_j$ contains taxon $x_i$. It is also natural to define the ``dual'' encoding. Given an $n \times m$ binary matrix $M$, the \emph{cluster encoding} of $M$ is a cluster set $Clus(M)$ containing a set of $m$ clusters $\{c_1, ..., c_m\}$ on taxon
set $\{x_1,...,x_n\}$ such that $c_j$ contains $x_i$ if and only $M_{i,j}$ has value 1. Clearly both encodings
can be produced in polynomial time.

The following result was presented in \cite{gusfieldNewton} and is to some extent implicit in \cite{kanghard} (and thus should be attributed to these two groups of authors) although to the best of our knowledge has never been formally written down. It shows that in a very strong sense the construction of phylogenetic networks from clusters, and recombination networks from binary characters under the all-0 root, unrestricted multiple crossover variant, are equivalent.

\begin{observation}
\label{obs:equivalence}
Given a cluster set $\mathcal{C}$, any phylogenetic network $N$ that represents $\mathcal{C}$ can be relabelled (after possibly a trivial modification) to obtain a
recombination network that represents $Mat(\mathcal{C})$ under the unrestricted multiple crossover variant with all-0 root. Given a binary matrix
$M$, any recombination network that represents $M$ under the unrestricted multiple crossover variant with all-0 root can be relabelled (after a possibly
trivial modification) to obtain a phylogenetic network that represents $Clus(M)$.
\end{observation}
\begin{proof}
The core idea is that the edges which represent clusters will become the edges upon which mutations from 0 to 1 will occur, and vice-versa. We will now formalise this.

Consider first a cluster set $\mathcal{C} = \{c_1, ..., c_m\}$ and a phylogenetic network $N$ that represents it. If necessary we first modify $N$ slightly to ensure that
every reticulation has outdegree exactly 1. Now, for each cluster $c_j \in \mathcal{C}$ there exists some tree $T_j$ on
$\mathcal{X}$ that is displayed by $N$ and which represents $c_j$. To obtain the recombination network for $Mat(\mathcal{C})$ we relabel as follows: the root of $N$ receives
the all-0 sequence and for each $c_j$ ($1 \leq j \leq m$) we locate the edge $e_j$ in $T_j$ that represents $c_j$, and fix some subdivision of $T_j$ in $N$. The edge $e_j$ will
thus correspond to a directed path of edges in $N$; we arbitrarily
choose one edge from this path as the edge at which character $j$ mutates from 0 to 1. (We can assume without loss of generality that this is not a reticulation edge). For each
node $v$ in $N$ we say that character $j$ has value 1 if and only if $v$ lies in the subdivision of $T_j$ that we fixed and the node $v'$ in $T_j$ to which it corresponds, is reachable
in $T_j$ from $e_j$ by a directed path. In particular, each character at a reticulation $v$ inherits its value from the node immediately preceding $v$ in the subdivision.

Given an $n \times m$ binary matrix $M$ and a recombination network $N$ that represents it under the unrestricted multiple crossover variant with all-0 root, we first
ensure that reticulations in $N$ with outdegree 0 are modified to have outdegree exactly 1. Now, we can relabel $N$ as
follows. The leaf labelled with row $i$ of $M$ is mapped to taxon $x_i$ of $\mathcal{X}$. Now, recall that the $j$th column of $M$ corresponds to cluster $c_j \in
Clus(M)$. Consider any such $j$. At every node $v$ in $N$ it is either (i) unambiguous from which parent of $v$ the value of character $j$ was
inherited, or (ii) it is ambiguous, in which case we can arbitrarily choose any such parent, or (iii) character $j$ mutates from a 0 to 1 on one of the
edges feeding into $v$, in which case choose that edge. This induces a tree which will be a subdivision of some tree $T_j$ on $\mathcal{X}$. Furthermore, $T_j$ represents $c_j$, and we are
done.
\qed
\end{proof}

\begin{figure}[h]
  \centering
  \includegraphics[scale=.5]{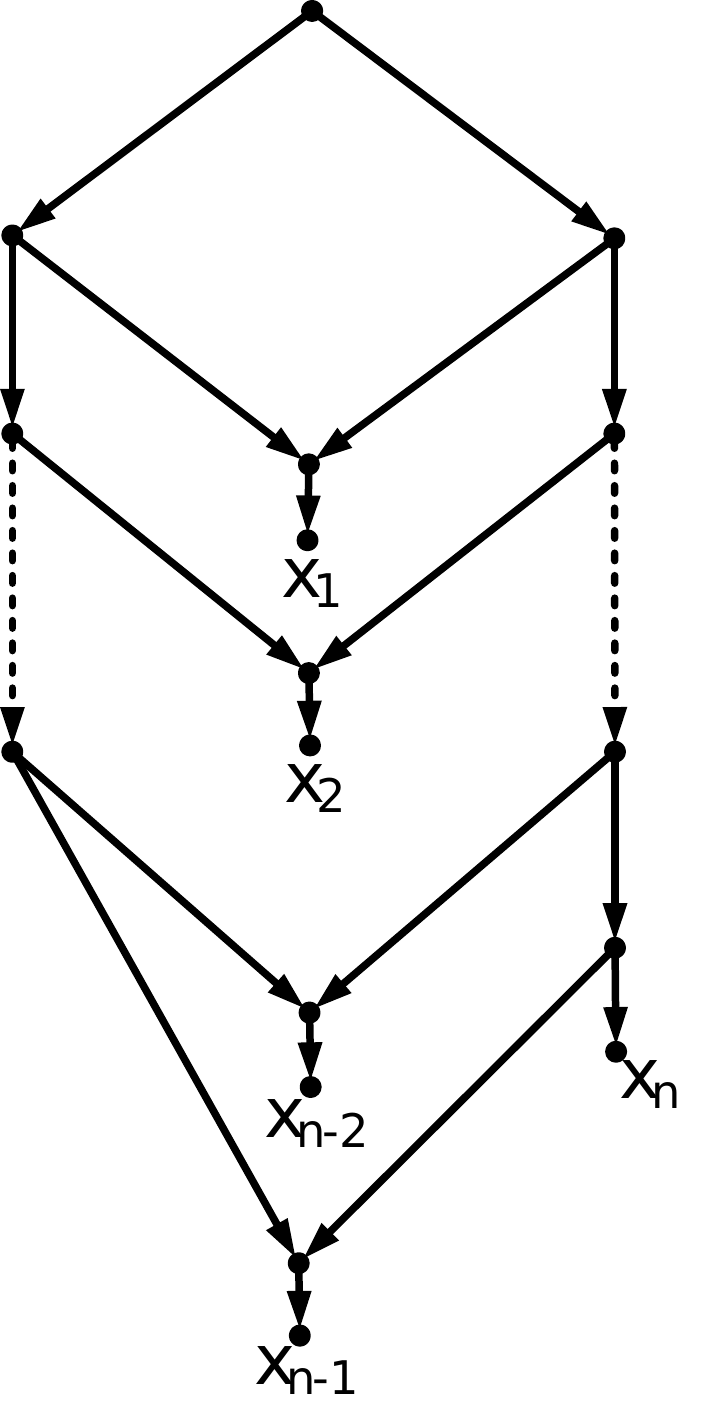}
  \caption{A network that is consistent with all $3\binom{n}{3}$ triplets and represents all $2^{n}-1$
clusters on taxon set $\mathcal{X} = \{x_1, ..., x_n\}$.}
  \label{fig:fulltripletset}
\end{figure}

\begin{corollary}
Given a cluster set $\mathcal{C}$, $r_{c}(\cC) = r_{uc}(Mat(\cC))$. Given a binary matrix $M$,
$r_{uc}(M) = r_{c}(Clus(M))$.
\end{corollary}
It is natural to wonder whether the single crossover variant is genuinely more restrictive than the unrestrained multiple crossover variant. Could it be, for example,
that the columns of an input matrix $M$ can always be re-ordered to obtain a matrix $M'$ such that $r_{sc}(M')  = r_{uc}(M)$? This is not so, as the following
simple example shows. We observe firstly that for a cluster set $\mathcal{C}$ on a set of taxa $\mathcal{X}$, $r_{c}(\mathcal{C}) \leq |\mathcal{X}|-1$. This follows because we can use
the
construction depicted in Figure \ref{fig:fulltripletset}. Now, for any integer $p \geq 5$ we let $\mathcal{C}_{p}$ be the set of all clusters that
contain exactly $\lfloor p/2 + 1 \rfloor$ elements of $\mathcal{X}$, where $\mathcal{X}$ is a taxon set on $p$ elements. Let $M = Mat(\mathcal{C}_p)$.
It follows by Observation \ref{obs:equivalence} that $r_{uc}(M) = r_{c}( Clus(M) ) = r_{c}( \mathcal{C}_p ) \leq p-1$.

Clearly $M$ has $k = \binom{p}{\lfloor p/2 + 1 \rfloor}$ columns and $k$ grows exponentially in $p$. Let $M'$ be obtained from $M$ by arbitrarily
permuting its columns. Note that any adjacent pair of columns in $M'$ fails the three-gamete test (with respect to the all-0 root) because two distinct clusters containing $\lfloor
p/2 + 1 \rfloor$ elements are necessarily incompatible. Hence, if we partition the columns of $M'$ into $\lfloor k/2 \rfloor$ disjoint pairs of adjacent columns, and apply a
composite haplotype bound (i.e. apply the haplotype bound
independently to
each disjoint pair of columns) \cite{song2}\cite{myers2003}, it follows that
$r_{sc}(M') \geq \lfloor k/2 \rfloor$. This lower bound grows exponentially in $p$, independently of the exact column permutation applied, while the upper bound on $r_{uc}(M)$ grows
only linearly. For $p \geq 5$ the gap between these bounds is already greater than zero.

We remark in passing that the ``root unknown'' version of the unrestrainted multiple crossover variant (let us denote this $r_{uc}^{*}$)
has an interesting interpretation when given $Mat(\cC)$ as input. In the ``root unknown'' version characters are allowed to start with value 1 at the root and
mutate at most once to 0 (as opposed to always starting with value 0 at the root and mutating at most once to 1). It follows then that
$r_{uc}^{*}(Mat(\cC))$ is the minimum number of reticulations ranging over all networks that, for each cluster $c \in \mathcal{C}$, represents $c$ \emph{or the
complementary cluster} $|\mathcal{X}|\setminus c$. It is easy to see that $r_{uc}^{*}(Mat(\mathcal{C}))$ can be significantly smaller than $r_{uc}(Mat(\mathcal{C}))$. For example,
consider the set $\mathcal{C}$ of all size-$2$ clusters on a size-3 taxon set $\mathcal{X}$. These clusters are mutually incompatible, so $r_{uc}(Mat(\mathcal{C}))
\geq 1$. However,the complement of each cluster is a singleton cluster, so (by choosing the all-1 root) $r_{uc}^{*}(Mat(\cC))=0$.

\subsection{Clusters and triplets coming from trees}\label{sec:tripclusterstrees}

Let us take a closer look at sets of triplets or clusters that are obtained from a set~$\cT$ of phylogenetic trees on the same set of taxa.
We will show that any phylogenetic network that represents~$Cl(\cT)$ is consistent
with~$Tr(\cT)$. It follows that representing all triplets requires at most as many reticulations as representing all clusters. Moreover, quite obviously, representing all clusters requires at
most as many reticulations as representing the trees themselves. Thus,

\begin{equation}\label{eq:tripletsclusterstrees}
r_{tr}(\cT) \leq r_{c}(\cT) \leq r_t(\cT) \enspace.
\end{equation}

Furthermore, this is true not only with respect to minimizing the number of reticulations, but with respect to minimizing any property of the networks, e.g. level:

\begin{equation}\label{eq:tripletsclusterstreeslevel}
\ell_{tr}(\cT) \leq \ell_{c}(\cT) \leq \ell_t(\cT) \enspace.
\end{equation}

We will show that each of the inequalities in~(\ref{eq:tripletsclusterstrees}) and~(\ref{eq:tripletsclusterstreeslevel}) is strict for some set of trees~$\cT$.

First, in order to prove~(\ref{eq:tripletsclusterstrees}) and~(\ref{eq:tripletsclusterstreeslevel}), we show an important relation between $Tr(\cT)$ and $Cl(\cT)$.

\begin{lemma}
For any three taxa~$x,y,z\in\cX$ holds that~$xy|z\in Tr(T)$ if and only if there exists a cluster~$C\in Cl(T)$ with~$x,y\in C$ and~$z\notin C$.
\label{lem:tripclus}
\end{lemma}
\begin{proof}
First suppose that there is a cluster~$C\in Cl(T)$ such that~$x,y\in C$ and $z\notin C$. Then the triplet~$xy|z$ is consistent with~$T$ and hence~$xy|z\in Tr(T)$.

Now suppose that $xy|z\in Tr(T)$. Then the triplet $xy|z$ is displayed by~$T$ and hence there is a subtree~$T'$ of~$T$ such that $xy|z$ can be obtained from~$T'$ by suppressing nodes with indegree one and outdegree one. This subtree~$T'$ contains exactly one node with indegree one and outdegree two. Let~$C$ be the set of taxa reachable from this node. Then, $x,y\in C$, $z\notin C$ and $C\in Cl(T)$.
\qed
\end{proof}

It follows that, for any set~$\cT$ of trees on the same set~$\cX$ of taxa, $Cl(\cT)$ uniquely determines $Tr(\cT)$.

We will now prove the following proposition, from which correctness of~(\ref{eq:tripletsclusterstrees}) and~(\ref{eq:tripletsclusterstreeslevel}) follows.

\begin{proposition}
For any set~$\cT$ of trees on the same set~$\cX$ of taxa, any phylogenetic network on~$\cX$ representing~$Cl(\cT)$ is consistent with~$Tr(\cT)$.
\end{proposition}
\begin{proof}
Let~$N$ be a phylogenetic network on~$\cX$ representing~$Cl(\cT)$. Consider a triplet $xy|z\in Tr(\cT)$.
By Lemma~\ref{lem:tripclus}, there is a cluster~$C\in Cl(T)$ (for some $T \in \mathcal{T}$) with~$x,y\in C$ and~$z\notin C$. Cluster~$C$ is represented by~$N$ (in the softwired 
sense) and hence there exists 
a phylogenetic tree~$T_C$ on~$\cX$ that is displayed by~$N$ and has~$C\in Cl(T_C)$. Because $x,y\in C$ and $z\notin C$, it follows that $xy|z$ is displayed by~$T_C$. Since~$T_C$ is displayed by~$N$, it follows that $xy|z$ is displayed by~$N$. Hence,~$N$ is consistent with $xy|z$.
\qed
\end{proof}

Before proceeding further, the following two lemmas will be of use throughout the rest of the article.

\begin{lemma}
\label{lem:binary}
Let $N$ be a phylogenetic network on $\mathcal{X}$. Then we can transform $N$ into a binary
phylogenetic network $N'$ such that $N'$ has the same reticulation number and level as $N$ and if $T$ is a binary tree displayed by $N$ then $T$ is also displayed by $N'$.
\end{lemma}
\begin{proof}
The transformation is very simple (and can clearly be conducted in polynomial time, if necessary). To begin with, each reticulation $v$ with outdegree 0 (which will be necessarily labelled
with some taxon $x \in \mathcal{X}$) is
transformed into a reticulation with outdegree 1 as follows. We introduce a new node $v'$, add the edge $(v,v')$ and move label $x$ to node $v'$.
Next we deal with nodes $v$ that have both indegree and outdegree greater than 1. Here we replace the node $v$ by an edge $(v_1, v_2)$ such that the edges incoming to $v$
now enter $v_1$, and the edges outgoing from $v$ now exit from $v_2$. Subsequently nodes with indegree at most 1, and outdegree $d\geq 3$, can
be replaced by a chain of $(d - 1)$ nodes of indegree at most 1 and outdegree 2. Nodes with indegree $d \geq 3$ and outdegree 1 can be
replaced by a chain of $(d -1)$ nodes of indegree 2 and outdegree 1. The critical observation is that if a binary tree $T$ is displayed by $N$ then there is a subdivision of
$T$ in $N$ which is also binary. This means that for each node $v$ in $N$ the subdivision uses at most two outgoing edges of $v$ and at most one incoming edge of $v$. Hence
the subdivision can easily be extended to become a subdivision within $N'$. \qed
\end{proof}

\begin{lemma}
\label{lem:binary2}
Let $N$ be a phylogenetic network on $\mathcal{X}$ and $\mathcal{T}$ a set of binary trees on $\mathcal{X}$. Then there exists a binary phylogenetic network $N'$ on $\mathcal{X}$
such that (a) $N'$ has the same reticulation number and level as $N$, (b) if $N$ displays all trees in $\mathcal{T}$ then so too does $N'$, (c) if $N$ is consistent with
$Tr(\mathcal{T})$ then so too is $N'$ and (d) if $N$ represents $Cl(\mathcal{T})$ then so too does $N'$.
\end{lemma}
\begin{proof}
(a) and (b) are immediate from Lemma \ref{lem:binary}. For (c) note that for each triplet $t \in Tr(\mathcal{T})$ there is some subdivision of $t$ in $N$. A triplet $t$ is binary,
and thus so too is any subdivision of $t$, so we can apply the same argument as used in Lemma \ref{lem:binary}. For (d), note that for each cluster $c \in Cl(\mathcal{T})$ there
is some tree $T$ on $\mathcal{X}$ which is displayed by $N$ and which represents $c$. $T$ is perhaps not binary, and thus a subdivision of it in $N$ is perhaps also not binary,
so after the transformation described in Lemma \ref{lem:binary} this subdivision will have become the subdivision of some binary tree $T'$. However, $T'$ is a refinement of $T$
i.e. $Cl(T) \subseteq Cl(T')$ so $c$ is also represented by $N'$. \qed
\end{proof}

\begin{figure}[h]
  \centering
   \includegraphics[scale=.5]{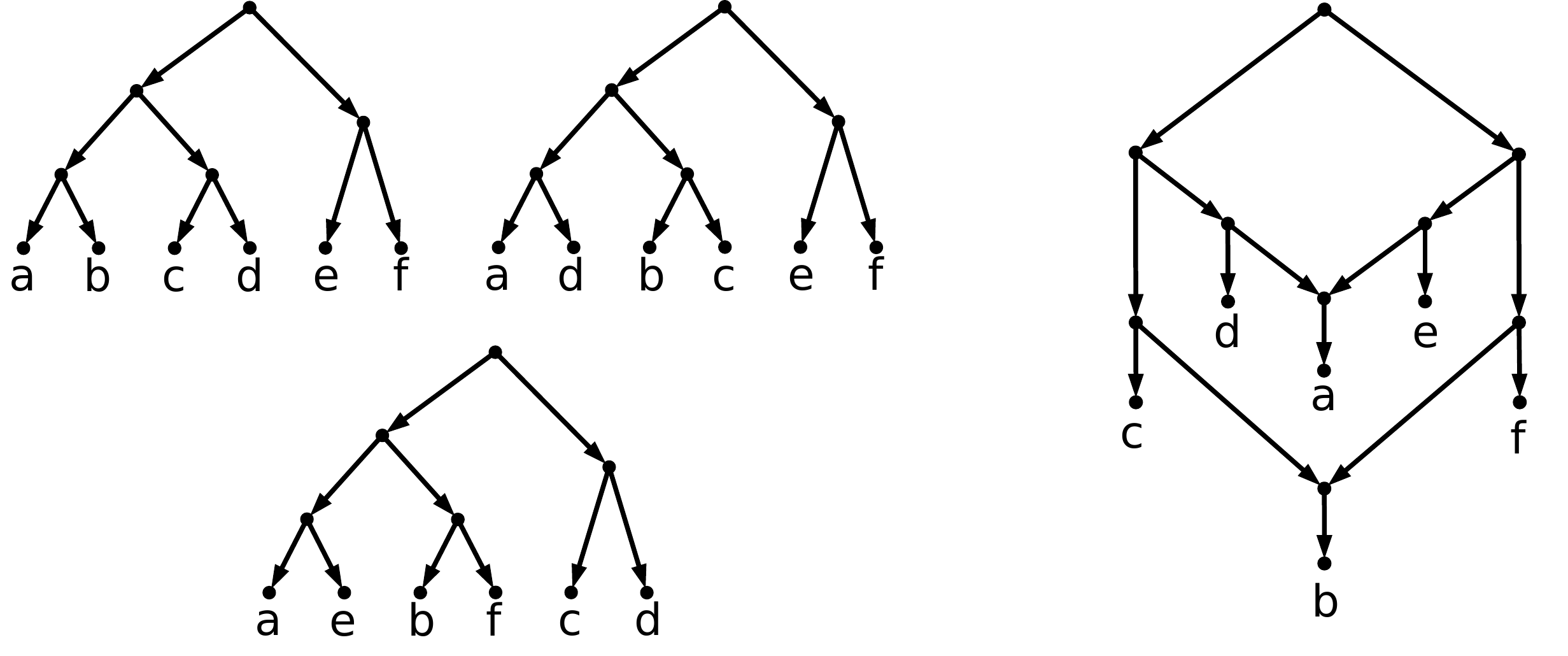}
  \caption{The triplets obtained from the three threes on the left are consistent with the level-2 network on the right containing two reticulations. However,
any network representing all the clusters from these trees will have at least three reticulations and be level-3 or higher.}
  \label{fig:ineq1tight}
\end{figure}

We will now show that each of the inequalities in~(\ref{eq:tripletsclusterstrees}) and~(\ref{eq:tripletsclusterstreeslevel}) is strict for some set of trees. To do so for the first inequality in each formula, consider
the set $\mathcal{T}$ of three trees, and the network $N$, shown in Figure~\ref{fig:ineq1tight}. It is easy to check that $N$ is consistent with all
the triplets in $Tr(\mathcal{T})$. However, any network that represents $Cl(\mathcal{T})$ requires at least 3 reticulations, and will be level-3 or higher, as can be verified by a straightforward (but technical) case analysis or by using the program \textsc{Cass} \cite{cassISMB}. Specifically: if a level-1 or level-2 network existed that represented $Cl(\mathcal{T})$ then \textsc{Cass} would definitely find it, and it does not.

\begin{figure}[h]
  \centering
  \includegraphics[scale=.5]{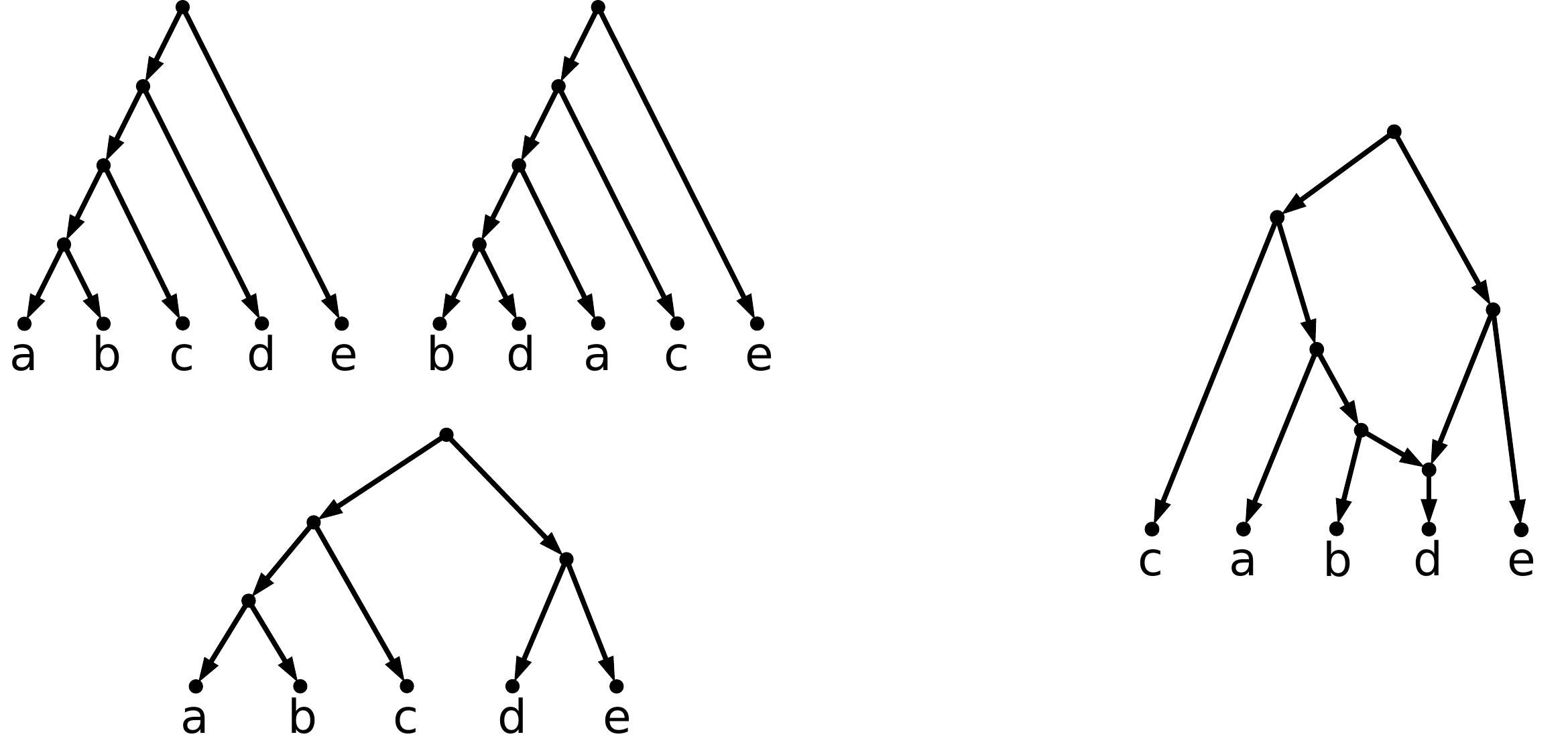}
  \caption{The level-1 network on the right with a single reticulation represents the union of the clusters (and triplets) obtained from the three trees on the left. However, any network
that displays all three trees will have at least two reticulations and have level at least two.}
  \label{fig:3tree}
\end{figure}

Figure~\ref{fig:3tree} shows a set~$\cT$ of trees for which the second inequality in~(\ref{eq:tripletsclusterstrees}) and~(\ref{eq:tripletsclusterstreeslevel}) is strict. A level-1
network with one reticulation is shown that represents all clusters from the three trees. However, a network with~$k$ reticulations can display at most $2^{k}$
distinct trees, so any network that displays all three trees will require at least two reticulations. It will also have level at least 2, because a (without loss of generality) binary level-1 network
displaying all three trees would have two nontrivial biconnected components, and thus all three trees would have a common non-singleton cluster, but this is not so.

Although we do not present a proof, empirical experiments furthermore suggest that it is possible to ``boost'' the example given in Figure~\ref{fig:3tree} to create sets of three binary trees $\mathcal{T}$ such that the gap between $r_{t}(\mathcal{T})$ and $r_{c}(\mathcal{T})$ can be made arbitrarily large
\cite{gapExperiment}.

\subsection{Clusters and triplets coming from two binary trees}
\label{subsec:2bt}

This section presents the main results of this paper. We will show that the number of reticulations necessary to represent the clusters from two binary trees on the same taxa is equal to the number of reticulations necessary to represent the trees themselves. In addition, we will show that also the number of reticulations necessary to represent all triplets from the two trees is equal to the number of reticulations necessary to represent the trees themselves. Moreover, we will show that the same is true when not the number of reticulations but the level of the networks is minimized. This means that for data coming from two binary trees on the same set of taxa, the tree-, cluster- and triplet problems all coincide.

Let~$\cT$ be a set containing two binary phylogenetic trees on the same set of taxa. Recall that~$Cl(\cT)$ is the set of all clusters from both trees in~$\cT$ and~$Tr(\cT)$ is the set of all triplets from both trees. We start by showing that the minimum number of reticulations in a network consistent with $Tr(\cT)$ is equal to the minimum number of reticulations in a network displaying both trees in~$\cT$. The fact that also the number of reticulations necessary to represent $Cl(\cT)$ is the same will be a corollary. After this corollary we will show that the results also hold for level-minimization.

First, however, some context is necessary. As mentioned earlier \cite{bordewich} fixed the partially correct result of \cite{wang} to prove that computation
of $r_{sc}$ is NP-hard. The correct part of the proof in \cite{wang}, Claim 2, essentially showed that,
for a set $\mathcal{T} = \{T_1, T_2\}$ of two binary trees on a set $\mathcal{X}$ of
taxa, $r_{t}(\mathcal{T}) \leq r_{uc}(M^{*})$ where $M^{*}$ is the concatenation
of $Mat(Clus(T_1))$ and $Mat(Clus(T_2))$ into a single matrix containing $4(n-1)$ columns (i.e. characters) and $|\mathcal{X}|$ rows. By (\ref{eq:scuc}) they thus also proved that
that $r_{t}(\mathcal{T}) \leq r_{sc}(M^{*})$ and this fact is used in \cite{bordewich}\footnote{The specific column ordering in $M^{*}$ - first the clusters from $T_1$ in
arbitrary order, and then the clusters from $T_2$ in arbitrary order - is important for establishing that $r_{t}(\mathcal{T}) \leq r_{sc}(M^{*})$. In particular, it is easy
to construct instances $\{T_1, T_2\}$ such that a bad permutation of the columns of $M^{*}$ causes $r_{sc}(M^{*})$ to be arbitrarily larger than $r_t(\mathcal{T})$.}.
Now, observe that $Clus(M^{*})$ is equal to
$Cl(\mathcal{T})$. Hence, by Observation
\ref{obs:equivalence}, $r_{t}(\mathcal{T}) \leq r_{uc}(M^{*}) = r_{c}(\mathcal{T})$. It is clear that $r_{c}(\mathcal{T}) \leq r_{t}(\mathcal{T})$ and hence
$r_{t}(\mathcal{T}) = r_{c}(\mathcal{T})$. In this sense the equivalence of $r_{t}(\mathcal{T})$ and $r_{c}(\mathcal{T})$ for pairs of binary trees was already implicitly
present in the literature. However, given (a) the lack of clarity in the proof of \cite{wang}, (b) the fact that Observation \ref{obs:equivalence}
has only been implicitly present in the literature up until now and (c) the desire to produce a unification result which also includes triplets, we have decided that it is useful to directly and explicitly prove this two-tree result and to explore its consequences.

\begin{theorem}\label{thm:twotrees}
If~$\cT = \{T_1, T_2\}$ consists of two binary phylogenetic trees on the same set of taxa, $r_{tr}(\cT)=r_t(\cT)$.
\end{theorem}
\begin{proof}

To increase the clarity of the proof we write $r_t(T_1, T_2)$ as shorthand for $r_t( \{ T_1, T_2 \} )$ and $r_{tr}(T_1, T_2)$
as shorthand for $r_{tr}(\{T_1, T_2\})$.

Clearly, $r_t(T_1, T_2)\geq r_{tr}(T_1, T_2)$, since any phylogenetic network displaying~$T_1$ and~$T_2$ is consistent with all triplets from~$T_1$ and~$T_2$. It remains to
show $r_t(T_1, T_2)\leq r_{tr}(T_1, T_2)$.

Suppose this is not true. Let~$n$ be the number of leaves in a smallest counter example, i.e.~$n$ is the smallest number such that there exist two binary phylogenetic trees $T_1$ and $T_2$
on a set of taxa~$\mathcal{X}$ with~$|\mathcal{X}|=n$ such that~$r_t(T_1, T_2)>r_{tr}(T_1, T_2)$. Clearly $n\geq 3$.
Let~$N_t$ be a phylogenetic network on~$\mathcal{X}$ with $r_t(T_1, T_2)$ reticulations that displays~$T_1$ and~$T_2$ and let~$N_{tr}$ be a phylogenetic network on~$\mathcal{X}$ with
$r_{tr}(T_1, T_2)$ reticulations that is consistent with all triplets in~$T_1$ and~$T_2$.

We may assume by Lemma \ref{lem:binary2} that~$N_{tr}$ and~$N_t$ are binary. We define a \emph{reticulation leaf} as a leaf whose parent is a reticulation and a \emph{cherry} as two leaves with a common parent.

We first prove that any binary phylogenetic network contains either a reticulation leaf or a cherry. Suppose that this is not true and let~$N$ be a smallest counter example, i.e. $N$ has no
reticulation leaves and no cherries and has a minimum number of leaves over all such networks. Take any leaf~$x$ of~$N$ and let~$p$ be its parent. It cannot be a reticulation, so~$p$ is
either a split node or the root. In both cases, we delete~$x$ and contract the remaining edge leaving~$p$, giving a smaller counter example. We conclude that any binary phylogenetic network
contains either a reticulation leaf or a cherry. Hence, this is also true for~$N_{tr}$.

First suppose that~$N_{tr}$ contains a cherry. Let this cherry consist of leaves~$a,b$ and their common parent~$v$. Then~$\{a,b\}$ is a cluster of~$T_1$ and of~$T_2$ i.e. they both
contain an edge whose set of leaf descendants is exactly $\{a,b\}$. If this was not so, then at least one of $T_1$ and $T_2$ would be consistent with a triplet $ac|b$ or $bc|a$ for some $c
\not \in \{a,b\}$ and such a triplet is not consistent with $N_{tr}$. It follows that each of~$T_1$ and~$T_2$ contains a cherry with leaves~$a,b$. Let~$T_1'$ and~$T_2'$ be the trees obtained
from~$T_1,T_2$ respectively by deleting leaves~$a$ and~$b$ and labeling their common
parent by a new label~$ab$. Now, Theorem~1 of Baroni et al.~\cite{BSS06} states that,
given a phylogenetic tree~$T$ and a cluster~$C\in Cl(T)$, let~$T|C$ denote the subtree of~$T$ on taxon set~$C$ and let~$T^{C\rightarrow c}$ denote the phylogenetic tree obtained from~$T$ by
replacing the subtree on~$C$ by a new leaf~$c$. Then, $r_t(T_1,T_2)=r_t(T_1|C,T_2|C)+r_t(T_1^{C\rightarrow c},T_2^{C\rightarrow c})$ whenever $C \in Cl(T_1) \cap Cl(T_2)$. Hence,
if we take $C = \{a,b\}$ we have that $r_t(T_1',T_2')=r_t(T_1,T_2)$.

Furthermore, $r_{tr}(T_1', T_2')\leq r_{tr}( T_1, T_2)$ because deleting~$a$ and~$b$ from~$N_{tr}$ and labelling~$v$ by~$ab$
leads to a phylogenetic network with $r_{tr}(T_1, T_2)$ reticulations that is consistent with all triplets in~$T_1'$ and~$T_2'$. We conclude that $$r_t(T_1',T_2')=r_t(T_1,T_2)>r_{tr}(T_1, T_2)\geq r_{tr}(T_1', T_2')\enspace .$$ Hence, we have constructed a smaller counter example, which shows a contradiction.

Now suppose that~$N_{tr}$ contains a reticulation leaf. Let~$x$ be such a leaf and~$r$ its parent. Let~$N_{tr}\backslash_x$ be the result of removing~$x$ and~$r$ from~$N_{tr}$. Let~$N_t\backslash_x$ be the result of removing~$x$ from~$N_t$ and removing the former parent of~$x$ as well if it is a reticulation. Let~$T_1\backslash_x$ and~$T_2\backslash_x$ be the trees
obtained from~$T_1$ and~$T_2$ respectively by removing~$x$ and contracting the remaining edge leaving the former parent of~$p$. That is, do the following for $i\in\{1,2\}$. Let~$p_i$ be the former parent of~$x$. If~$p_i$ is not the root, there is one edge~$(u^x_i,p_i)$ entering~$p_i$ and one edge~$(p_i,v^x_i)$ leaving~$p_i$. Remove~$p_i$ and replace the edges~$(u^x_i,p_i)$,$(p_i,v^x_i)$ by a single edge~$(u^x_i,v^x_i)$. We will use the edges $(u^x_i,v^x_i)$ later on. If~$p_i$ is the root, we remove~$x$ and~$p_i$ and leave $(u^x_i,v^x_i)$ undefined.

First observe that~$N_{tr}\backslash_x$ is consistent with all triplets of~$T_1\backslash_x$ and~$T_2\backslash_x$. Moreover, since~$N_{tr}\backslash_x$ contains one reticulation fewer than~$N_{tr}$,
\begin{equation}\label{eq:first}
r_{tr}(T_1\backslash_x, T_2\backslash_x) < r_{tr}(T_1, T_2) < r_t(T_1,T_2)
\end{equation}
and hence
\[
r_{tr}(T_1\backslash_x, T_2\backslash_x) \leq r_t(T_1,T_2) - 2\enspace .
\]

Now observe that~$N_t\backslash_x$ displays~$T_1\backslash_x$ and~$T_2\backslash_x$. We will show that
\begin{equation}\label{eq:second}
r_t(T_1\backslash_x,T_2\backslash_x) \geq r_t(T_1,T_2) - 1 \enspace .
\end{equation}
Together,~(\ref{eq:first}) and~(\ref{eq:second}) imply that
\[
r_{tr}(T_1\backslash_x, T_2\backslash_x) \leq r_t(T_1,T_2) - 2 \leq r_t(T_1\backslash_x,T_2\backslash_x) - 1
\]
and hence that we have obtained a smaller counter example, which is a contradiction.

It remains to prove~(\ref{eq:second}). Let~$N'$ be a phylogenetic network on~$\mathcal{X}\setminus\{x\}$ with $r_t(T_1\backslash_x,T_2\backslash_x)$ reticulations that
displays~$T_1\backslash_x$ and~$T_2\backslash_x$. Since~$T_1\backslash_x$ is displayed by~$N'$, there exists a subgraph~$E_1$ of~$N'$ that is a subdivision of~$T_1\backslash_x$ (an embedding
of~$T_1\backslash_x$ into~$N'$). Similarly, let~$E_2$ be a subgraph of~$N'$ that is a subdivision of~$T_2\backslash_x$. We will now use the edges $(u^x_1,v^x_1)$ and $(u^x_2,v^x_2)$ that we
introduced when defining $T_1\backslash_x$ and $T_2\backslash_x$. For~$i\in\{1,2\}$, if the edge~$(u^x_i,v^x_i)$ has been defined, we define the edge~$e_i$ as follows. The edge
$(u^x_i,v^x_i)$ corresponds to a directed path in~$E_i$. Let~$e_i$ be any edge of this path. Notice that~$e_i$ is an edge of~$N'$.

Let~$N^+$ be the network obtained by subdividing~$e_1$ and~$e_2$ and making~$x$ a reticulation leaf below the new nodes. To be precise, for $i\in\{1,2\}$, if~$e_i=(u_i,v_i)$ has been
defined, replace~$e_i$ by $(u_i,n_i),(n_i,v_i)$ with~$n_i$ a new node. If $(u_i,v_i)$ has not been defined, add a new root~$n_i$ and an edge from~$n_i$ to the old root. Finally, add a leaf
labelled~$x$, a new reticulation~$r$ and edges $(n_1,r),(n_2,r)$ and $(r,x)$.

Observe that~$N^+$ displays~$T_1$ and~$T_2$, because we can simply extend each of the embeddings~$E_1$ and~$E_2$ by the new edges leading to the leaf~$x$. Moreover,~$N^+$ contains exactly one
reticulation more than~$N'$. Thus, $r_t(T_1,T_2)\leq r_t(T_1\backslash_x,T_2\backslash_x) + 1$, which remained to be shown.
\qed
\end{proof}

\begin{corollary}
If~$\mathcal{T}$ consists of two binary phylogenetic trees on the same set of taxa,
$$r_{tr}(\mathcal{T}) = r_{c}(\mathcal{T}) = r_t(\mathcal{T})\enspace .$$
\label{cor:RETtripsclusterstrees}
\end{corollary}
\begin{proof}
Follows from combining Theorem~\ref{thm:twotrees} with~(\ref{eq:tripletsclusterstrees}).
\qed
\end{proof}

\begin{figure}[h]
  \centering
  \includegraphics[scale=.5]{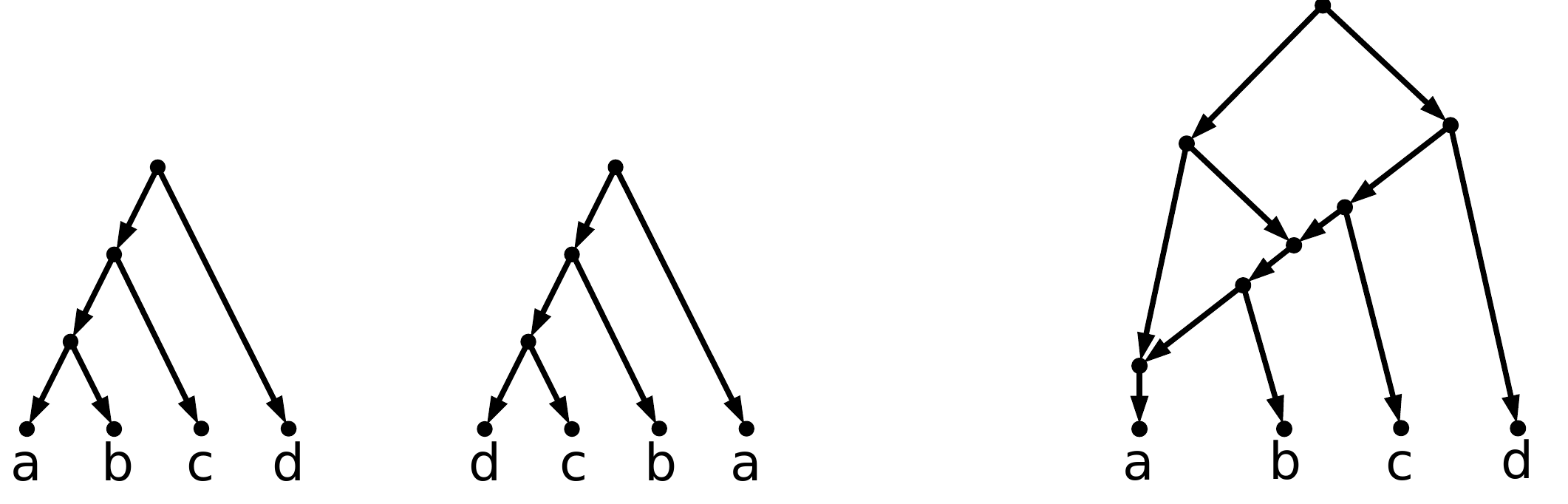}
  \caption{The network on the right represents the union of the clusters (and triplets) obtained from the two trees on the left, but it does not display both trees.}
  \label{fig:notalwaystrees}
\end{figure}

Given this result it is natural to ask whether every network that represents all the clusters (or triplets) from two binary trees $T_1$ and $T_2$ on the same taxon set,
and having a minimum number of reticulations, also displays $T_1$ and $T_2$. This is not so. Consider the two trees in Figure \ref{fig:notalwaystrees}. It is
easy to check that two reticulations are necessary and sufficient to display both these trees. The network in this figure contains two reticulations and represents the union
of the clusters (and triplets) from both trees, but it does not display both trees.

\begin{figure}[h]
  \centering
  \includegraphics[scale=.5]{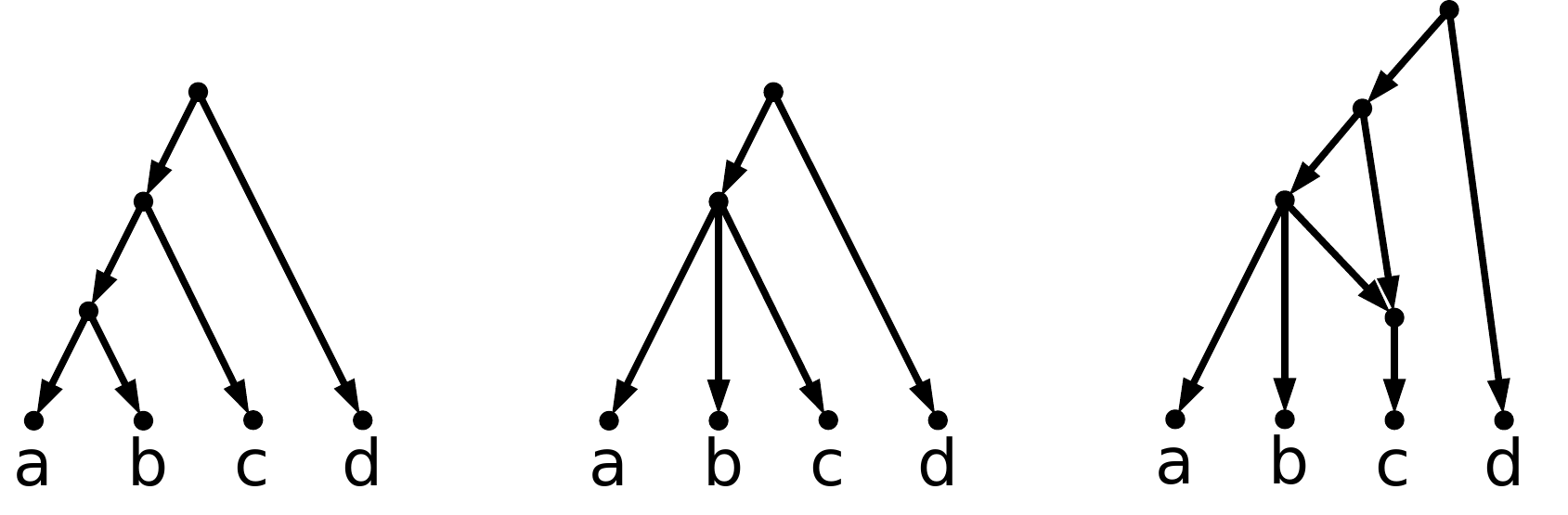}
  \caption{The network on the right displays the two trees on the left: at least one reticulation is necessary. However, the tree on the left is sufficient to represent the union of the
clusters (or triplets) obtained from both trees.}
  \label{fig:nonbinary}
\end{figure}

We note that Theorem~\ref{thm:twotrees} and Corollary \ref{cor:RETtripsclusterstrees} do not hold for sets of three or more trees, as demonstrated in Section \ref{sec:tripclusterstrees} by Figure~\ref{fig:3tree}. In addition,
they also do not hold for two possibly non-binary trees, as demonstrated by Figure~\ref{fig:nonbinary}.

For a binary phylogenetic network $N$ on $\mathcal{X}$ the notion of a \emph{cut-edge} is well-defined: an edge $(u,v)$ whose removal disconnects $N$. A cut-edge is \emph{trivial} if
at least one of the disconnected components created by its removal contains fewer than 2 taxa from $\mathcal{X}$, and is called \emph{nontrivial} otherwise. $N$ is said to be \emph{simple}
if it does not contain any nontrivial cut-edges.

\begin{theorem}
If~$\mathcal{T}$ consists of two binary phylogenetic trees on the same set of taxa,
$$\ell_{tr}(\cT) = \ell_{c}(\cT) = \ell_t(\cT)\enspace .$$
\label{thm:LEVtripsclusterstrees}
\end{theorem}
\begin{proof}
By~(\ref{eq:tripletsclusterstreeslevel}), it suffices to show $\ell_t(\cT) \leq \ell_{tr}(\cT)$. We do so by induction on~$|\cX|$. The base case for~$|\cX|\leq 2$ is clear. Now consider a set
of trees~$\cT$ on~$\cX$ with~$|\cX|=n$. Let~$N_{t}$ be a network that displays all trees in~$\cT$ and has optimal level~$\ell_t(\cT)$. Similarly, let~$N_{tr}$ be a network consistent
with~$Tr(\cT)$ that has optimal level~$\ell_{tr}(\cT)$. By Lemma \ref{lem:binary2} we may assume that $N_t$ and $N_{tr}$ are both binary. We distinguish three cases.

First suppose that neither~$N_t$ nor~$N_{tr}$ contains nontrivial cut-edges, i.e. that~$N_t$ is a simple level-$\ell_t(\cT)$ network and~$N_{tr}$ is a simple level-$\ell_{tr}(\cT)$
network. In that case, the number of reticulations in~$N_t$ is equal to $\ell_t(\cT)$. So, $r_t(\cT)\leq\ell_t(\cT)$. At the same time, $r_t(\cT)\geq\ell_t(\cT)$, since the number of
reticulations in any network is at least equal to its level. Thus, $r_t(\cT)=\ell_t(\cT)$. Similarly, $r_{tr}(\cT)=\ell_{tr}(\cT)$. Moreover, by Theorem~\ref{thm:twotrees},
$r_{tr}(\cT)=r_t(\cT)$ and we can conclude that $\ell_{tr}(\cT) = r_{tr}(\cT)= r_t(\cT) =\ell_t(\cT)$.

Now suppose that~$N_t$ contains at least one nontrivial cut-edge and let~$e$ be such an edge. Let~$C$ be the set of taxa reachable from~$e$ by a directed path. Let~$\cT|C$ be the set of trees
obtained by restricting each of the trees in~$\cT$ to the taxa in~$C$ and let~$\cT^{C\rightarrow c}$ denote the set of trees obtained by collapsing, in each tree in~$\cT$, the subtree on~$C$
by a single leaf labelled~$c$. We claim that

\begin{align*}
\ell_t(\cT) & \leq \max\{ \ell_t(\cT|C) , \ell_t(\cT^{C\rightarrow c}) \}\\
& = \max\{ \ell_{tr}(\cT|C) , \ell_{tr}(\cT^{C\rightarrow c}) \}\\
& \leq \ell_{tr}(\cT)\enspace .
\end{align*}

To see that $\ell_t(\cT) \leq \max\{ \ell_t(\cT|C) , \ell_t(\cT^{C\rightarrow c}) \}$, notice that any network displaying~$\cT^{C\rightarrow c}$ can be combined with any network
displaying~$\cT|C$ in order to obtain a network displaying~$\cT$. This can be done by replacing the leaf~$c$ of the network displaying~$\cT^{C\rightarrow c}$ by the network
displaying~$\cT|C$. The network obtained in this way displays~$\cT$ and its level is equal to the maximum of the levels of the networks displaying~$\cT^{C\rightarrow c}$ and~$\cT|C$. So,
$\ell_t(\cT) \leq \max\{ \ell_t(\cT|C) , \ell_t(\cT^{C\rightarrow c}) \}$. Then we use that $\ell_t(\cT|C) = \ell_{tr}(\cT|C)$ and $\ell_t(\cT^{C\rightarrow c}) =
\ell_{tr}(\cT^{C\rightarrow c})$ by induction. To prove the last inequality, observe that $\ell_{tr}(\cT|C)\leq \ell_{tr}(\cT)$ because removing leaves can not increase the level.
In addition, $\ell_{tr}(\cT^{C\rightarrow c})\leq \ell_{tr}(\cT)$ because $\cT^{C\rightarrow c}$ can be constructed by removing all leaves in~$C$ except for one, which is
relabeled~$c$, and removing or relabeling leaves can not increase the level.

The final case is that~$N_{tr}$ contains a nontrivial cut-edge~$e$. Let~$C$ be the set of taxa that can be reached from~$e$ by a directed path in~$N_{tr}$. Clearly, for~$x,y\in C$
and~$z\notin C$, $xy|z\in Tr(\cT)$. Thus,~$C$ is a cluster of each of the trees of~$\cT$. Therefore, we can argue in the same way as in the previous case that $\ell_t(\cT) \leq
\ell_{tr}(\cT)$.
\qed
\end{proof}

\section{Complexity Consequences}
\label{sec:complexity}

Theorem \ref{thm:twotrees} and Corollary \ref{cor:RETtripsclusterstrees} allow us to elegantly settle several complexity questions in the phylogenetic network literature that have been open for some time, and to significantly strengthen some already existing hardness results.

\begin{corollary}\label{cor:REThard}
Computing $r_c(\cT)$ and computing $r_{tr}(\cT)$ are both NP-hard and APX-hard, even for sets~$\cT$ consisting of two binary trees on the same set of taxa.
\end{corollary}
\begin{proof}
Follows from Corollary~\ref{cor:RETtripsclusterstrees} and the fact that computing~$r_t(\cT)$, for sets~$\cT$ consisting of two binary trees on the same set of taxa, is NP-hard and APX-hard~\cite{bordewich}.
\qed
\end{proof}

It follows directly that the following two problems are NP-hard and APX-hard.

\medskip
\noindent\begin{tabular}{lp{0.85\textwidth}}
\multicolumn{2}{l}{\textsc{MinRetClusters}} \\
\textit{Instance:} & A set~$\cX$ of taxa and a set~$\cC$ of clusters on~$\cX$.\\
\textit{Objective:} & Construct a phylogenetic network on~$\cX$ that represents each cluster in~$\cC$ and has a minimum number of reticulations over all such networks.\\
\end{tabular}

\medskip
\noindent\begin{tabular}{lp{0.85\textwidth}}
\multicolumn{2}{l}{\textsc{MinRetTriplets}} \\
\textit{Instance:} & A set~$\cX$ of taxa and a set~$\cR$ of triplets on~$\cX$.\\
\textit{Objective:} & Construct a phylogenetic network on~$\cX$ that is consistent with each triplet in~$\cC$ and has a minimum number of reticulations over all such networks.\\
\end{tabular}

Moreover, the latter problem is even NP-hard and APX-hard for dense sets of triplets. This strengthens a result by Jansson et al.~\cite{JanssonEtAl2006}, who showed that \textsc{MinRetTriplets} and \textsc{MinLevTriplets} are NP-hard, by constructing a non-dense set of triplets such that positive instances of the NP-complete problem \textsc{Set Splitting} corresponded to a level-1 network with exactly one reticulation. Corollary~\ref{cor:REThard} extends this result by showing that \textsc{MinRetTriplets} is even NP-hard for dense sets of triplets and that it is hard to approximate (APX-hard).

We now turn our attention to the problems that minimize level.

\begin{theorem}\label{thm:mltIsHard}
Computing $\ell_t(\cT)$ is NP-hard and APX-hard, even for sets~$\cT$ consisting of two binary trees on the same set of taxa.
\end{theorem}
\begin{proof}
We again reduce from the problem of computing~$r_t(\cT)$, for sets~$\cT$ consisting of two binary trees on the same set of taxa. We first reduce this problem to the restriction to pairs of trees~$T_1,T_2$ that do not have a
common non-singleton cluster. Call this restricted problem \textsc{ResMinRetTrees}.

Consider a set~$\cT$ consisting of two binary phylogenetic trees~$T_1,T_2$ on a set~$\cX$ of taxa. Recall Theorem~1 of Baroni et al.~\cite{BSS06} and the application of it described in the proof of Theorem~\ref{thm:twotrees} in
this article. To summarise, $r_t(T_1,T_2)=r_t(T_1|C,T_2|C)+r_t(T_1^{C\rightarrow c},T_2^{C\rightarrow c})$ whenever $C \in Cl(T_1) \cap Cl(T_2)$. Thus, repeatedly applying the Baroni theorem, we obtain a collection of at most
polynomially-many instances of \textsc{ResMinRetTrees} such that the minimum reticulation number of the original instance is equal to the sum of the minimum reticulation numbers of the obtained instances of \textsc{ResMinRetTrees}. Thus, we can solve the original instance by solving each instance of \textsc{ResMinRetTrees}. This completes the reduction.

We continue by reducing \textsc{ResMinRetTrees} to the problem of computing $\ell_t(\cT)$. Consider an instance $(\mathcal{X},T_1,T_2)$ of \textsc{ResMinRetTrees}. Let $\mathcal{T}=\{T_1, T_2\}$. We will prove that $\ell_{t}(\mathcal{T})$ = $r_{t}(\mathcal{T})$ and this will complete the reduction. Clearly $\ell_{t}(\mathcal{T}) \leq r_{t}(\mathcal{T})$. Suppose then for the sake of contradiction that $\ell_{t}(\mathcal{T}) < r_{t}(\mathcal{T})$. If that is the case, then any level-$\ell_t(\cT)$ network that displays~$T_1$ and~$T_2$ contains at least two nontrivial biconnected components. By Lemma~\ref{lem:binary2}, there exists a binary such phylogenetic network~$N$. Since this network contains at least two nontrivial biconnected components, it contains a cut-edge~$e=(u,v)$ such that at least two taxa are reachable from~$v$ (by a directed path) and at least one taxon is not. Define cluster~$E$ to contain all taxa that are reachable from~$v$ in~$N$. Thus, $|E|\geq 2$. $T_1$ and~$T_2$ are both displayed by~$N$ so, for $i \in \{1,2\}$, there is a subdivision of~$T_i$ in~$N$. Fix any such subdivision. So, each edge of $T_i$ maps to a directed path of one or more edges in~$N$. Both subdivisions must pass through $(u,v)$ and it thus
follows that~$E$ is a non-singleton cluster of both~$T_1$ and~$T_2$, giving us a contradiction. This completes the NP-hardness proof.

To see that computing $\ell_t(\cT)$ is not only NP-hard but also APX-hard, observe that
\textsc{ResMinRetTrees} is APX-hard because (as shown above) $r_t(\cT)$ can be computed by simply adding up the
optima of polynomially-many instances of \textsc{ResMinRetTrees}. This additivity means that an $\epsilon$-approximation to \textsc{ResMinRetTrees} yields an $\epsilon$-approximation for the problem of computing $r_t(\cT)$. Combining this with the optimality-preserving reduction from \textsc{ResMinRetTrees} to the problem of computing $\ell_t(\cT)$ described above gives the desired result.
\qed
\end{proof}

It follows directly that the following problem is NP-hard and APX-hard.

\medskip
\noindent\begin{tabular}{lp{0.85\textwidth}}
\multicolumn{2}{l}{\textsc{MinLevTrees}} \\
\textit{Instance:} & A set~$\cX$ of taxa and a set~$\cT$ of phylogenetic trees on~$\cX$.\\
\textit{Objective:} & Construct a level-$k$ phylogenetic network on~$\cX$ that displays each tree in~$\cT$ and such that~$k$ is as small as possible.\\
\end{tabular}

\begin{corollary}
Computing $\ell_c(\cT)$ and computing $\ell_{tr}(\cT)$ are both NP-hard and APX-hard, even for sets~$\cT$ consisting of two binary trees on the same set of taxa.
\label{cor:clusLevAPX}
\end{corollary}
\begin{proof}
Follows from Theorem~\ref{thm:LEVtripsclusterstrees} and Theorem~\ref{thm:mltIsHard}. \qed
\end{proof}

Thus, also the following two problems are NP-hard and APX-hard.

\medskip
\noindent\begin{tabular}{lp{0.85\textwidth}}
\multicolumn{2}{l}{\textsc{MinLevClusters}} \\
\textit{Instance:} & A set~$\cX$ of taxa and a set~$\cC$ of clusters on~$\cX$.\\
\textit{Objective:} & Construct a level-$k$ phylogenetic network on~$\cX$ that represents each cluster in~$\cC$ and such that~$k$ is as small as possible.\\
\end{tabular}

\medskip
\noindent\begin{tabular}{lp{0.85\textwidth}}
\multicolumn{2}{l}{\textsc{MinLevTriplets}} \\
\textit{Instance:} & A set~$\cX$ of taxa and a set~$\cR$ of triplets on~$\cX$.\\
\textit{Objective:} & Construct a level-$k$ phylogenetic network on~$\cX$ that is consistent with each triplet in~$\cR$ and such that~$k$ is as small as possible.\\
\end{tabular}

Moreover, the latter problem is even NP-hard and APX-hard for dense sets of triplets.

\section{Concluding Remarks}

In this article, we have proven an important unification result that shows that when computing the minimum number of reticulations (or minimum level) required to represent data obtained from two binary trees on the same taxon set, it does not matter whether one calculates this using trees, triplets or clusters. In the process of proving this, we have clarified a number of confusing issues in the literature.

The unification result has the interesting practical consequence that the two-tree case thus forms an interesting benchmark for comparing the performance of different phylogenetic network software. It was already empirically observed in \cite{cassISMB}, for example, that for a specific two-tree data set the independently developed programs \textsc{Cass} (which takes clusters as input, and attempts to minimise level), \textsc{PIRN} (which takes trees as input, and attempts to minimise the reticulation number) and \textsc{HybridInterleave} (which takes two binary trees as input, and minimises the reticulation number) all returned the same optimum. The intriguing possibility thus exists of creating hybrid software for the two-tree problem by combining the best parts of several existing software packages. It should be noted, however, that the \emph{networks} achieving these optima are not always transferrable. For example, a network obtaining the minimum number of reticulations under the cluster model does not automatically display both the trees.

It is also interesting to view our results next to other two-tree findings in the literature. Phillips and Warnow~\cite{phillipswarnow96} showed that, given a set of clusters coming from two trees, it is polynomial-time solvable to find a phylogenetic tree consistent with a maximum number of clusters, while this problem is NP-hard for three or more trees. Another interesting two-tree result was discovered by Bordewich, Semple and Spillner~\cite{bordewichsemplespillner2009}. They found a polynomial-time algorithm for finding an optimal set of taxa that maximizes the weighted sum of the phylogenetic diversity across two phylogenetic trees, while also this problem is NP-hard for three or more trees. It would be interesting to try and identify general families of objective functions (i.e. optimization criteria) for which the two-tree case is special.

On the other hand, we have shown that the tree, triplet and cluster models already start to diverge for three binary trees on the same set of taxa.
A natural follow-up question is thus: can we predict under what circumstances the models significantly differ, and what does it say about our choice of model if sometimes one model requires significantly more reticulations, or higher level, than another? The ``triplet $\leq$ cluster $\leq$ trees'' inequality from Section \ref{sec:tripclusterstrees} suggests that in appropriate combinations existing software for triplets, clusters and trees could be used to develop lower and upper bounds for each other, but under what circumstances are these bounds strong?

\bibliography{twotrees_v23}

\end{document}